\def   \ni {\noindent}

\def   \ssk {\vskip  5truept}

\def   \bsk {\vskip 15truept}
 
\def   \newpage {\vfill\eject}
\def   \newline {\hfil\break}

\documentstyle[epsfig]{article}
\begin{document}

\hsize 5truein
\vsize 8truein
\font\abstract=cmr8
\font\keywords=cmr8
\font\caption=cmr8
\font\references=cmr8
\font\text=cmr10
\font\affiliation=cmssi10
\font\author=cmss10
\font\mc=cmss8
\font\title=cmssbx10 scaled\magstep2
\font\alcit=cmti7 scaled\magstephalf
\font\alcin=cmr6 
\font\ita=cmti8
\font\mma=cmr8
\def\ref{\par\noindent\hangindent 15pt}
\null

                
\title{\ni GALACTIC SUPERLUMINAL SOURCES}                                               

\bsk \bsk
\author{\ni B.~A.~Harmon}                                                       
\bsk
\affiliation{ES84, NASA/Marshall Space Flight Center, Huntsville, AL, USA
35812}                                                
\bsk
\baselineskip = 12pt

\abstract{ABSTRACT \ni
A new class of X-ray sources was clearly established with the discovery of
highly relativistic radio jets from the galactic sources GRS 1915+105 and
GRO J1655-40.  Both of these objects have given us a broader view of black
holes and the formation of jets, yet they also show the complexity of the
accretion environment near relativistic objects. The fast apparent motion of
the jets, their luminosity and variability, high energy spectrum, and
approximate scaling to the behavior of active galactic nuclei, certainly
warrant the description ``microquasar".  I present a review of the observational
data on these sources, and discuss where we stand on a physical picture of 
GRS 1915+105 and GRO J1655-40 as taken from multi-wavelength studies. I also 
point out other galactic sources which share some of the properties of the
microquasars, and what to look for as a high energy ``signature" in future
observations. 
}                                                    
\bsk
\baselineskip = 12pt

\bsk
\baselineskip = 12pt


\text{\ni 1. INTRODUCTION
\ssk
\ni     

    A new class of galactic X-ray transient sources was identified
with the discovery of superluminal radio jets in GRS 1915+105 
\cite{Mira94} and GRO J1655-40\cite{Tin95, Hjell95}.  
Recently, thanks to rapid followup radio observations, other jet sources,
although not superluminal, have been observed. No evidence of X-ray bursts or
coherent pulses have been found in any of these objects, and are therefore 
likely to be black hole candidates (BHC) (the mass of GRO J1655-40 has 
been measured to be about 7M$_{\odot}$ \cite{Oro97}. A more recent analysis 
yields 4.1 $<$ M$_{x}$ $<$ 6.6M$_{\odot}$  (90\% conf. limit) \cite{Ph98}).

Because of the physical similarity of the galactic
superluminal sources to their extragalactic counterparts, they have
also been called ``microquasars".  An analogy drawn 
between jets in active galactic nuclei (AGN) and accreting binaries 
shows consistent scaling in luminosity, size and variability \cite{Sams96}. 
The advantage in studying the galactic sources comes from their size and
proximity; the timescale of
the variations are a factor of 10$^{6}$ shorter, and the modest
Doppler boosting
usually allows both the approaching and receding jets to be seen.   

Here
we review the experimental evidence for a relationship between wavelength bands
from gamma rays to
radio, with emphasis on possible 
temporal or spectral signatures in
the high energy data for jet formation.  Then, building on
the framework established for AGN jets, we discuss the
theoretical implications of these observations and the role of
the disk magnetic field.

\begin{figure}
\centerline{\psfig{file=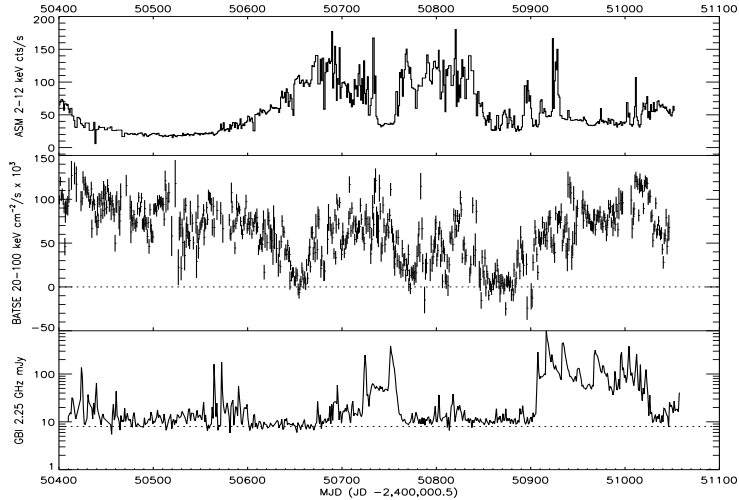, 
height=7cm, width=10cm}}
\caption{FIGURE 1. Recent history of GRS 1915+105 from (top to bottom)
the RXTE ASM (2-12 keV), CGRO BATSE (20-100 keV), and GBI
(2.25 GHz) 50400 = 11/13/96, 51100 = 10/14/98.}
\end{figure}


\bsk
\ni 2. LONG TERM BEHAVIOR
\ssk
\ni 
GRS 1915+105 was first detected by GRANAT/WATCH in
X-rays in 1992 \cite{Cas92}.
Numerous outbursts have
occurred since then, and it is likely that GRS 1915+105 has not returned to 
quiescence since
that time. The radio counterpart 
was
monitored sparsely until late 1993 \cite{Rod95}, when it
was observed as  a strong transient radio source ($\sim$Jy).  Prior to this, the
only galactic sources
known to have large radio flares were the persistent sources SS 433, Cir X-1,
and Cyg X-3.  
The observed radio spectrum is consistent with
electron synchrotron
radiation, varying between flat (at lower flux levels) to
and characteristically steep (higher levels) during flares.  The synchrotron
emission can extend at times into the infrared \cite{Sams96,Eik98}. 
In Fig. 1,
we show monitoring results for GRS 1915+105 from 1996-1998 in the radio band
from the Green Bank 
Interferometer (GBI), soft X-rays measured with the {\it Rossi X-ray Timing 
Explorer}
(RXTE) All-Sky Monitor (ASM)  and 
hard X-rays with the {\it Compton Gamma Ray Observatory} (CGRO)
Burst and Transient Source Experiment (BATSE).    
We see that radio flares occur in association with 
hard X-ray/gamma-ray ($>$20 keV) outbursts.  A general increase at 2.25 GHz can be
seen in the 1996-98 interval.  During outburst the flux density can
reach 50 mJy and remain in a plateau-like state for several weeks
\cite{Fo96}, with occasional flares reaching 1 Jy.  The soft X-ray flux
(2-12 keV) has distinct quiescent and flaring substates which appear
largely independent of the other bands on timescales of days or more.
There is a tendency for the hard X-rays to be high during the quiescent
soft X-ray states, but no direct one-to-one correlation is seen between
the two bands.

GRO J1655-40 was discovered with BATSE in 1994 \cite{Zh94}.  It too 
had several outbursts in the high energy regime; it has since returned to
quiescence by 1997.  Radio outbursts to several Jy were seen in conjunction 
with early oubursts in 1994-95, and the radio counterpart reappeared briefly
in a 1996 outburst.  It is the only superluminal source with a
an optically-determined mass function and it also has an F6-type 
stellar companion \cite{Oro97}.

\bsk
\ni 3. X-RAY SIGNATURE OF PLASMA EJECTION
\ssk
\ni 
3.1 Temporal Behavior
\ssk
\ni
The excitement
of discovery of GRS 1915+105 and GRO J1655-40 have now stimulated 
a large amount of new
and valuable data on the superluminal sources.
High energy observations have revealed \cite{Harm95,
Harm97} evidence that accretion is 
definitely related to the formation of jets.  
In 1994 (see Fig. 2), a strong hard X-ray outburst of GRS 1915+105
was accompanied by 
plasmoid ejections
\cite{Mira94} which
indicated for the first time that superluminal motion was observed in a galactic
source.
\begin{figure}
\centerline{\psfig{file=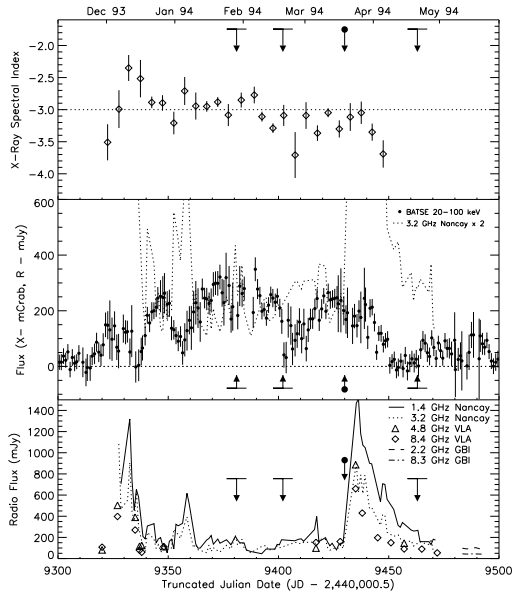,width=7cm}}
\caption{FIGURE 2. The 1994 outburst of GRS 1915+105 in the hard
X-ray (BATSE) and GHz radio band (VLA, Nancay and GBI).  The top frame
shows the photon spectral index of the X-ray emission, the middle
frame, the X-ray intensity and 3.2 GHz radio flux (Nancay), and bottom:
several radio bandpasses. Approximate times of radio
ejection events are indicated by arrows. Figure from ref. \cite{Harm97}.}
\end{figure}
The tendency for the hard X-ray emission to dip or decrease
during the 
largest flares ($>$200 mJy) can be seen in the middle frame of Fig. 2.
This was the first and only time that direct correlations
with the large flares were seen.  During more recent hard X-ray 
outbursts as in Fig. 1, we continued to observe enhanced radio
emission, but at variable intensities and generally lower than intense,
longer flares shown in Fig. 2.
The apparent difference in X-ray behavior between the high radio flux levels 
in 1994 and at times when the radio flux is lower may be due to a lack
of sensitivity at high energy to the smaller/shorter ejection events.  

X-ray observations with the PCA on RXTE have revealed cyclic (or
approximately repeating) intensity variations consisting of flares and
dips
lasting a few thousand seconds in GRS 1915+105 \cite{Mor97, Bel97}. 
The intensity variations are accompanied by dramatic changes in 
spectral hardness.   
There is also a clear, delayed response in the infrared 
and radio bands \cite{Eik98,Mira98}.} characteristic
of synchrotron radiation. This behavior is not present in all observations, 
and was only observed in GRS 1915+105.  Occasional dips in the intensity
of other sources such as GRO J1655-40 and 4U 1630-47 lasting several minutes
appear to be an absorption phenomenon \cite{Kuu98}.

\begin{figure}
\centerline{\psfig{file=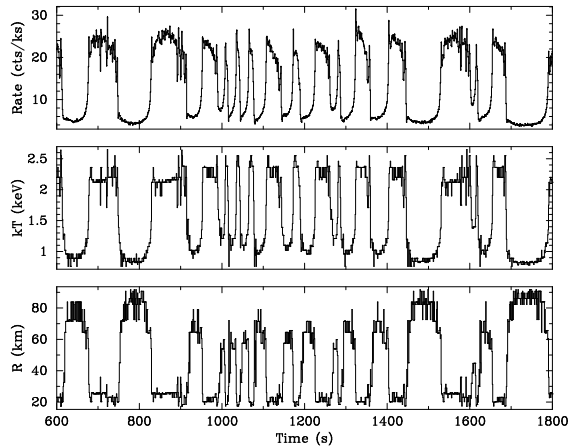, 
width=7cm,angle=-90}}
\caption{FIGURE 3. PCA data for GRS 1915+105 showing the count rate,
and derived radius and temperature parameters as a function of
time. Fig. from ref. \cite{Bel97}.}
\end{figure}

Belloni et al. \cite{Bel97} 
have demonstrated that the X-ray data of Fig. 3 can be ordered according to a 
derived value
of the inner disk radius {\it R$_{in}$}, which scales according to the length 
of the
troughs and the intensity of flares.  They 
describe the cycle as an emptying and refilling of accretion material from the 
inner disk region.  Although the radio flux during the flares and dips
is not sufficient to prove that a jet is present, if 
material is being removed via a ``mini-jet", it
could represent a scaled-down version of the larger
events observed in Fig. 2.

\bsk
\ni
3.2 Spectral Behavior
\ssk
\ni
The spectral behavior of the superluminal sources, much less BHC
in all their various accretion states, is not well understood.  We do know 
that the X-ray/gamma ray range is dominated by multi-color black-body
and/or Comptonization
components as in other BHC \cite{Zh97a}.  
The jet sources tend to exhibit a highly variable disk black-body 
component (from which {\it R$_{in}$} was
obtained for GRS 1915+105), and a high energy tail which has a high
degree of independence from the soft component.

In addition, two other spectral properties are shared between
the superluminal sources.  Both GRS 1915+105 and GRO J1655-40 spectra
show complicated
Fe line profiles \cite{Ueda98}.  To interpret them as
due to a single reflection, absorption, or disk-line component is
difficult, but Ueda et al.
\cite{Ueda98} suggest that the features near 7 keV seen in 
Japanese satellite ASCA data are due to 
absorption by 
He-like and H-like iron ions in an anisotropic, hot plasma.
Such a plasma may be similar to that forming the hot ion torus in AGN.
The second shared property of
the superluminal sources is the steep,
power law spectral shape in the hard X-ray/gamma ray regime (see Fig. 4).
Photon number indices in the 20-200 keV band range from $-$2.5 to $-$3.5, 
with spectral slopes that 
do not seem to change drastically or appear to be strongly correlated with
luminosity as in other black hole candidates. OSSE has observed emission
from GRO J1655-40 to 700 keV with no evidence of a cutoff
\cite{Grov98}.  This suggests
a nonthermal origin for the emission; however, better statistics are
required for confirmation.
\begin{figure}
\centerline{\psfig{file=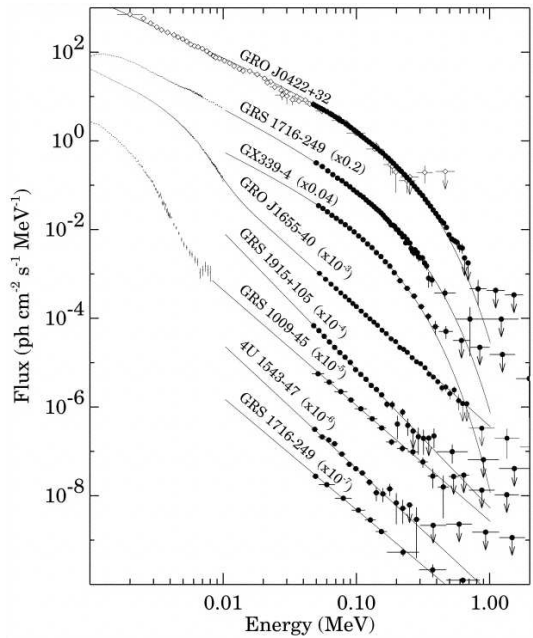, 
width=7cm}}
\caption{FIGURE 4. OSSE spectra for various black hole candidates.
Figure from ref. \cite{Grov98}}.
\end{figure}

The broadband spectral shape of the superluminal sources
is usually associated with the high or ultrahigh states of BHC where
the mass accretion rate may reach near-Eddington luminosities.  
The steep power law shape can result from
Comptonization of a nonthermal electron distribution such as might
be encountered
inside the last stable orbit (radial bulk motion) around the black hole
\cite{Chak95} 
This model has been used successfully \cite{Shra98} to fit spectra for 
GRO J1655-40 and GRS 1915+105.  The Comptonization model must have a high
energy cutoff, and cannot explain emission above a few hundred keV.
We point out that a number of BHC, which apparently do not exhibit jets,
also have this high energy spectral shape, e.g., Nova Muscae, GRS 1009-45,
and 4U 1543-47.  (See Fig. 4).  This may be an observational bias, due to
incompleteness in historical coverage of their outbursts in
radio/infrared bands.    

\bsk
\ni 4. THEORETICAL CONSIDERATIONS 
\ssk
\ni 
The discovery rate of black hole transients per year is now about 5-6
from the nearly continuous coverage of all-sky monitors such as BATSE 
and the ASM.  Two of the
more recently discovered transients, CI Cam and XTE J1550-564, 
have bright ($\sim$100 mJy) radio counterparts. 
Very long baseline interferometry indicates
that the radiation comes from clouds of radio-emitting plasma associated
with jets.  
It is therefore possible that jets are common among
black hole transients, although the persistent outbursting and jet
production in GRS 1915+105 may be rare.  

Most black hole transients, such as GRO J1655-40, tend to be active
in the GHz band over the first few weeks or months of the outburst, if at all.
This associates
relatively
large accretion rates with jet formation, typical of the
high to very high states.
There is also evidence that particular substates produce
jets, as in GRS 1915+105, when the source is highly variable, and undergoing
drastic changes in the broadband spectrum.  Rapid changes 
in the inner disk structure may alternately produce a highly luminous disk
in soft X-rays, or a fast outflow of high energy particles.  For now, we
do not know the origin of such substates.  Likely explanations overlap with
those of quasi-periodic oscillations in the accretion disk, which are 
beyond the scope of this paper (see contributions by Stella
and van der Klis, this meeting). 

Considerable theoretical work on the formation
and collimation of jets has come from observations of AGN over
many years.  It is thought that a magnetic field is generated within the 
nucleus by ordered motion of accreting gas and dust.
The disk magnetic field is likely to play a strong role in launching 
material out of the disk as well as collimating the outflow \cite{Blan82}.
This model, in its various forms, is attractive because
magnetically-driven jets scale easily to any size system from binaries to AGN.
Recent work with three dimensional
magnetohydrodynamical simulations provides an
explanation of the transient nature of the jets via instabilities in magnetic
pressure dominated (low $\beta$) disks \cite{Mat97} or magnetic ``switches" 
from the variations in coronal particle density \cite{Me97}.

Currently there is no
way to directly measure the strength of such magnetic fields near
galactic black holes.  Such fields are inferred from observations of
young stellar objects and the central regions of quasars. A weakness of 
magnetic field models
is that they provide no explanation for the
presence of jets in only some systems.  Various other observations in the 
optical
and infrared suggest the presence of massive companions with
high mass loss rates (such as in a brief period of evolution) that results
in conditions that are conducive to jet production, similar to Cyg X-3
and SS 433.  

It has also been proposed, as in AGN, that the spin of the black hole
may be a critical factor in jet production.  The angular momentum of the
black hole causes the event horizon to shrink, and thus the inner
edge of the accretion disk can extend closer to the black hole.  This
can increase soft X-ray production significantly.  Zhang et al. 
\cite{Zh97b}  have
proposed that the superluminal sources may contain black holes with angular
momentum near the maximum allowed value.  This
may account for the large soft X-ray emission seen at times
in the broadband spectra of the GRO J1655-40 and GRS 1915+105.  It
is not clear how the angular momentum of the black hole is extracted
by the jet;
a possible mechanism for this was given by Blandford and Znajek \cite{Blan77}. 
This scenario
provides an explanation for other black holes (such as Cyg X-1 in the 
hard/low
state and GS 2023+25) of not having prominent jets consistent with the lack of 
strong ultrasoft components and 
the steep power law tail extending into the gamma ray regime.

\bsk
\ni 5. THE FUTURE 
\ssk
\ni 

We hope to continue investigations of the known superluminal sources,
as well as new sources which may become active in the future.  
Clearly,
understanding has come by combining data from all parts of the electromagnetic
spectrum.  New instrumentation available soon
should allow us to probe both the ion species present in accretion disks
and perhaps even the composition of the jets themselves.  The high
energy part of the spectrum, which can be studied with both space-based
and ground-based instruments, can provide more clues about the acceleration
mechanism and the particle populations near black holes.  

\bsk
\baselineskip = 12pt
{\abstract \ni ACKNOWLEDGMENTS

BAH thanks C. Robinson for assistance with the preparation of BATSE
Earth occultation light curves.  This research has made use of data obtained
through the High Energy Astrophysics Science Archive Research Center Online
Service, provided by the NASA/Goddard Space Flight Center.  Data have
also been obtained through the Green Bank Interferometer (GBI) NASA monitoring
program. GBI is a facility of the National Science
Foundation operated by the National Radio Astronomy Observatory in support
of NASA High Energy Astrophysics programs.}

\bsk
\baselineskip = 12pt
\newpage
\ssk

\end{document}